\documentclass[12pt,amsfonts]{article}
\usepackage{amsfonts}
\def\ts{\textstyle}
\def\t{\textstyle}        
\def\half{{\textstyle{\frac{1}{2}}}}

\def\quarter{\textstyle{\frac{1}{4}}}

\def\H{{\cal H}}

\def\p{\phi}

\def\H{{\cal H}}

\def\v{\vskip.3em}

\def\ra{\rightarrow}

\def\tint{{\textstyle\int}}

\def\s{\hskip.08em}
\def\d{\partial}

\def\b{\begin{eqnarray*}}  
\def\e{\end{eqnarray*}}    
\def\bn{\begin{eqnarray}}  
\def\en{\end{eqnarray}}   
\def\<{\langle}
\def\>{\rangle}

\def\no{\nonumber}

\def\k{\kappa}

\def\{{\lbrace}
\def\}{\rbrace}

\title{Divergences in Scalar Quantum \\Field Theory: The Cause and the Cure}
\author{John R. Klauder\footnote{Email: klauder@phys.ufl.edu}\\
Department of Physics and\\Department of Mathematics\\
University of Florida\\
Gainesville, FL 32611-8440}
\date{ }

\begin{document}
\maketitle
\begin{abstract}
Covariant scalar fields exhibit divergences when quantized in two or
more spacetime dimensions: $n\ge2$.
Does  perturbation theory, effective theories, the renormalization group, etc., tell
us all there is to know about these problems? An
alternative approach identifies the cause of divergences as due to
the effort to multiplicatively relate two measures that are mutually
 singular,
while the cure for those divergences is to introduce an
$O(\hbar)$ counterterm that converts mutually singular measures into equivalent measures.
This procedure leads to a nontrivial, divergence-free formulation for
all $n\ge2$. Finally, a critical comparison of the new methods
with traditional procedures  is given.
\end{abstract}

\section{Introduction}
As a representative example of the models considered, we focus on a scalar field $\p(x)$, $x\in{\mathbb R}^n$, with the
classical (but imaginary time) action functional
\bn I(\p)=\tint\{\half[(\nabla\p)(x)^2+m_0^2\s\p(x)^2]+g_0\s\p(x)^4\s\}\,d^n\!x\;. \label{e1}\en
In turn, the quantization of this field may be addressed by the formal functional integral
  \bn S(h)\equiv{\cal M}\int e^{ (1/\hbar)\tint h(x)\s\p(x)\,d^n\!x-(1/\hbar)[I(\p)+ F(\p,\hbar)/2]}\;\Pi_x \s d\p(x)\;,\label{e2}\en
  where ${\cal M}$ is chosen so that $S(0)=1$, $h(x)$ is a smooth source function, and $F(\p,\hbar)$ represents an as-yet unspecified counterterm to control divergences, which should formally vanish as $\hbar\ra0$ so that the proper classical limit formally emerges. Rather than adopt a standard version of the counterterm, we seek a counterterm that eliminates all divergences.
  Although the formal
  functional integral (\ref{e2}) is essentially undefined, it can be given meaning by first introducing a lattice regularization in which the spacetime continuum is replaced with a periodic, hypercubic lattice with lattice spacing $a>0$ and with  $L<\infty$ sites along each axis. The sites themselves are labeled by multi-integers $k=\{k_0,k_1,\ldots,k_s\}\in{\mathbb Z}^n$, $n=s+1$, and $h_k$ and $\p_k$ denote field values at the $k$th lattice site; in particular, $k_0$ is designated as the Euclidean time variable. This regularization results in the $L^n$-dimensional integral
  \bn &&S_{latt}(h)\equiv M\int e^{ (1/\hbar)Z^{-{1/2}}\Sigma_k h_k\s\p_k\,a^n-(1/2\hbar)\s[\Sigma_{k,\s k^*}(\p_{k^*}-\p_k)^2\,a^{n-2}+\s m_0^2\s\Sigma_k\p_k^2\,a^n]}\no\\
  &&\hskip8em \times\,e^{-(1/\hbar)\s g_0\Sigma_k\s\p_k^4\,a^n-(1/2\s\hbar)\s\Sigma_k F_k(\p,\hbar)\,a^n}\;\Pi_k\s d\p_k\;.
  \label{e4} \en
  Here we have introduced the field-strength renormalization constant $Z$. The factors $Z$, $m_0^2$, and $g_0$ are
  treated as bare parameters implicitly dependent on the lattice spacing $a$,
 ${k^*}$ denotes one of the $n$ nearest neighbors in the positive direction from the site $k$, and $M$ is chosen so that $S_{latt}(0)=1$. The counterterm $F_k(\p,\hbar)$ also implicitly depends on $a$, and
  the notation $F_k(\p,\hbar)$ means that the formal, locally generated counterterm $F(\p,\hbar)$ may, when lattice regularized,  depend on finitely-many field values located within a small, finite region of the lattice around the site $k$; this issue will become clarified  when $F_k(\p,\hbar)$ is determined.

  Since the lattice regulation has led to finitely many integrations in (\ref{e4}), it is instructive to focus on the emergence of divergences as the continuum limit is taken, which we define as $a\ra0$, $L\ra\infty$, with  $a\s L$ fixed and finite. Divergences already arise  as $L\ra\infty$ without the need for $a\s L\ra\infty$ as well; for a discussion of the limit $a\s L\ra\infty$ see \cite{IOP}.

 \subsection{Simple examples of divergences and their removal}
  As an illustration of the kind of divergences of interest, let us initially consider an elementary example of moments, for $p\ge1$,  of a normalized probability distribution given, for $J<\infty$,  by
    \bn I_p\equiv\int [\s\Sigma_{j=1}^J y^2_j\s]^p\,e^{-\Sigma_{j=1}^J y_j^2}\;\Pi_{j=1}^J\s (dy_j/\sqrt{\pi})=O(J^p)\;.\label{e3}\en
    Here, the  approximate evaluation arises just by expanding the term $[\Sigma_{j=1}^J y^2_j]^p$ in the integrand, which leads to $J^p$ terms each of $O(1)$.  For any $p\ge1$, these moments all diverge as $J\ra\infty$. We normally attribute this kind of divergence to having an infinite number of
    integration variables, and we regularize similar integrals by restricting the number of integration variables (at least effectively, if not explicitly). Despite this natural assignment of the cause of divergences, it is in fact not true, in general, that an infinite number of integration variables is the cause of divergences. To illustrate that point, let us first change the integration variables to what we call ``hyperspherical coordinates'' defined by $y_j\equiv \k\s\eta_j$, $\k^2\equiv\Sigma_{j=1}^J\s y_j^2$, and $1\equiv\Sigma_{j=1}^J\s\eta_j^2$, where $0\le\k<\infty$ and $-1\le\eta_j\le1$. Expressed in the new integration variables, (\ref{e3}) becomes
       \bn I_p=\int \k^{2p}\,e^{-\k^2}\,\k^{J-1}\s d\k\;2\s\delta(1-\Sigma_{j=1}^J\eta_j^2)\,\Pi_{j=1}^J\s (d\eta_j/\sqrt{\pi})\;. \label{e5}\en
   No longer do we have $J^p$ terms each being $O(1)$. Instead, the incipient divergence as $J\ra\infty$ arises from a steepest descent evaluation of the integral over $\k$ in which we find that $\k^2=O(J)$, and thus the integral is $O(J^p)$ as it
   must be. Observe that the divergence arises from the unlimited growth of the power $J-1$ of the hyperspherical radius variable $\k$. Hypothetically, if we
    could change the $\k$-measure factor from $\k^{J-1}$ to $\k^{R-1}$, where $R>0$ is fixed and finite, and we also rescale the overall expression by $T_J$ (chosen so that the new distribution is normalized  for $p=0$ for all $J$), we are led to the expression
     \bn I'_p\equiv T_J\int \k^{2p}\,e^{-\k^2}\,\k^{R-1}\s d\k\;2\s\delta(1-\Sigma_{j=1}^J\eta_j^2)\,\Pi_{j=1}^J\s (d\eta_j/\sqrt{\pi})\; \label{e6}\en
   {\it for which all divergences disappear as $J\ra\infty$ without reducing the number of integration variables}. We call this procedure of changing the $\k$-measure factor: {\it measure mashing}.

   Why is it that divergences arise for (\ref{e5}) when they do not arise for (\ref{e6})? The answer lies in the {\it support of the measures}. To appreciate this remark, let us start with a very simple example. Consider the one-dimensional integral
      \bn \tint_c^{1+c} e^{i\s t\s u}\,du=e^{i\s t\s c}\,(\s e^{i\s t\s}-1)/(i\s t)\;.\en
    Apart from the phase factor, the integrand is the function  $Q_c(u)$ which is unity within the interval $[c,1+c]$ and zero elsewhere.
   Clearly, there exists no well-behaved function $W(u)$ such that $Q_{c'}(u)=e^{W(u)}\s Q_c(u)$ when
    $c'\not= c$.
    By contrast, if we {\it regularize} both $Q_{c'}(u)$ and $Q_c(u)$ by smoothing them out so that they both have the same support (e.g., the whole real line), there would indeed be a well-behaved function $W_{reg}(u)$ that would connect them. However, as the regularization is being removed, and the two supports tend toward their original values, the function $W_{reg}(u)$ will develop divergences for the simple reason, e.g., that it has to create something where there was nothing before. The lesson this example shows is:
   a perturbation analysis works when the beginning and ending measures have {\it equivalent support} and fails when the beginning and ending measures have (at least partially) {\it disjoint support}.

   Not all cases of (partially) disjoint support between measures are self evident as was the case for our one-dimensional example. The probability measure $d\mu_\beta(y)\equiv \Pi_{j=1}^\infty e^{-\beta y_j^2}\,\sqrt{\beta/\pi}\,dy_j$ on ${\mathbb R}^\infty$ is disjoint for unequal $\beta$ values as follows from the fact that, for
   $J<\infty$, the random variable $Y_J\equiv
   J^{-1}\Sigma_{j=1}^J \s e^{i\s t\s y_j}$ has the property that $\lim_{J\ra\infty}\,Y_J$ $=e^{-t^2/4\beta}$
    with probability one for all $t$. On expansion to order $t^2$, it follows that
     \bn \lim_{J\ra\infty} J^{-1}{\t\sum}_{j=1}^J\,y_j^2=1/2\beta \en
     for all sequences $\{y_j\}_{j=1}^\infty$ supported by $\mu_\beta$. Change $\beta$ and the old and new supports are disjoint! That fact alone is responsible for the divergence of the moments $I_p$, $p\ge1$, in (\ref{e3}).

     \section{Determining the proper counterterm for \\covariant models}
     We now return to the the lattice-regularized functional integral (\ref{e4}). In order for
     this mathematical expression to be physically relevant following a Wick rotation to real time, we impose the requirement of {\it reflection positivity}
     \cite{reflec}, which is assured if the counterterm satisfies
       \bn \Sigma_k\s F_k(\p,\hbar)\equiv\Sigma_{k_0}\s \{\Sigma'_k { F}_k(\p,\hbar)\} \en
       where each element $\{\Sigma'_k { F}_k(\p,\hbar)\}$ involves fields all of which have the same temporal value $k_0$, but may involve several other fields at nearby sites to $k$ in spatial directions only. (The primed sum $\Sigma'_k$ denotes a sum over a single spatial slice all sites of which have the same $k_0$.)

       Let us next consider (\ref{e4}) limited to a
       test function supported on a single spatial slice, e.g., $h_k\equiv a^{-1}\s \delta_{k_0,0}\, f_k$, for which (\ref{e4})
       becomes equivalent  \cite{IOP} to
       \bn S'_{latt}(f)=\int e^{Z^{-1/2}\Sigma'_k f_k\p_k\s a^s/\hbar}\,\Psi_0(\p)^2\,\Pi'_kd\p_k\;, \label{e10}\en
       where $\Psi_0(\p)$ denotes the normalized ground-state wave function of the Hamiltonian operator $\H$ for the problem, expressed in the Schr\"odinger representation, with
       the property that $\H\Psi_0(\p)=0$. (The primed product $\Pi'_k$ runs over all sites on a single
       spatial slice with a fixed value of Euclidean time $k_0$.) Expressed in hyperspherical coordinates for the $N'\equiv L^s$ sites on a spatial slice---for which $\p_k\equiv \k\s\eta_k$, $\k^2\equiv \Sigma'_k\p_k^2$, $1\equiv\Sigma'_k\eta_k^2$, $0\le\k<\infty$, and $-1\le\eta_k\le1$---(\ref{e10})
       becomes
         \bn S'_{latt}(f)=\int e^{\k\s Z^{-1/2}\Sigma'_kf_k\s\eta_k\,a^s/\hbar}\,\Psi_0(\k\s\eta)^2\,\k^{N'-1}\s d\k\,
         2\s\delta(1-\Sigma'_k\eta_k^2)\,\Pi'_k\s d\eta_k\;.\en
         In the continuum limit, as $L\ra\infty$ and therefore $N'=L^s\ra\infty$, divergences will generally
         arise because, in that limit, if parameters like $m_0$ or $g_0$, are changed, the measures
         become mutually singular due to the overwhelming influence of $N'$ in the measure factor
         $\k^{N'-1}$. However, just as in the example studied before, we can avoid that conclusion provided the ground-state distribution $\Psi_0(\p)^2$ contains a factor that serves to mash the measure.
         Specifically, we want the ground-state wave function to have the form
           \bn \Psi_0(\p)\hskip-1.4em&&= ~^{``} M'\,e^{-U(\p,\hbar,a)/2}\,\k^{-(N'-R)/2}\s^{\,"}\no\\
           &&=M' e^{-U(\k\eta,\hbar,a)/2}\,\k^{-(N'-R)/2}\Pi'_k[\Sigma'_l\s J_{k,l}\s\eta^2_l]^{-(1-R/N')/4}\no\\
               &&=M' e^{-U(\p,\hbar,a)/2}\,\Pi'_k[\Sigma'_l\s J_{k,l}\s\p^2_l]^{-(1-R/N')/4}\;.\en
               The first line (in quotes) indicates the qualitative $\k$-behavior that will effectively mash the measure, while the second and third lines illustrate a specific functional dependence on field variables that leads to the
               desired factor. Here we choose the constant coefficients $J_{k,l}\equiv1/(2s+1)$ for $l=k$ and for $l$ equal to each of the $2s$ spatially nearest neighbors to the site $k$; otherwise, $J_{k,l}\equiv0$. Thus, $\Sigma'_lJ_{k,l}\s\p_l^2$ provides an
               {\it average of field-squared values at and nearby the site $k$}. As part of the ground-state distribution, this factor is dominant for {\it small-field} values, and its form is no less fundamental than the rest of the ground-state distribution that is determined by the gradient, mass, and interaction terms that fix the {\it large-field} behavior. The factor $R/N'$ appears in the local expression of the small-field factor, and on physical grounds that quotient should not depend on the number of lattice sites in a spatial slice nor on the specific parameters mentioned above that define the model. Therefore, we can assume that $R\propto N'$, and so we set
                              \bn   R\equiv 2\s b\s a^s\s N'\;, \en
               where $b>0$ is a fixed factor with dimensions (Length)$^{-s}$ to make $R$ dimensionless. Even though the ground-state distribution diverges when certain of the $\eta_k$-factors  are simultaneously zero, these are all integrable singularities since whenever any subset of the $\{\eta_k\}$ variables are near zero, there are always fewer zero factors arising from the singularities thanks to the local averaging procedure; this very fact has motivated the averaging procedure.

               The choice we have made to mash the measure is not unique, but it is ``minimal'' in the sense that any other function of $\k$ would require dimensional factors.

               To obtain the required functional form of the ground-state wave function for small-field values, we choose our counterterm to build that feature into the Hamiltonian. In particular, the counterterm is a specific potential term of the form
               \bn\half \Sigma'_k\s{ F}_k(\p,\hbar)\,a^s\equiv \half\hbar^2\Sigma'_k{\cal F}_k(\p)\,a^s\en
               where, with $T(\p)\equiv \Pi'_r[\Sigma'_l J_{r,l}\s\p_l^2]^{-(1-2ba^s)/4}$,
                \bn {\cal F}_k(\p)\hskip-1,3em&& \equiv\frac{ a^{-2s}}{ T(\p)}\frac{\d^2 T(\p)}{\d\p_k^2}\no\\
                &&=\quarter\s(1-2ba^s)^2\s
          a^{-2s}\s\bigg({\ts\sum'_{\s t}}\s\frac{\t
  J_{t,\s k}\s \p_k}{\t[\Sigma'_m\s
  J_{t,\s m}\s\p_m^2]}\bigg)^2\no\\
  &&\hskip2em-\half\s(1-2ba^s)
  \s a^{-2s}\s{\ts\sum'_{\s t}}\s\frac{J_{t,\s k}}{[\Sigma'_m\s
  J_{t,\s m}\s\p^2_m]} \no\\
  &&\hskip2em+(1-2ba^s)
  \s a^{-2s}\s{\ts\sum'_{\s t}}\s\frac{J_{t,\s k}^2\s\p_k^2}{[\Sigma'_m\s
  J_{t,\s m}\s\p^2_m]^2}\;. \label{eF} \en
  Although ${\cal F}_k(\p)$ does not depend only on $\p_k$, it nevertheless becomes a local potential
  in the formal continuum limit.

  More generally, the full Hamiltonian operator including the desired counterterm is defined as
    \bn \H\hskip-1.3em&&=-\half\hbar^2\s a^{-2s}{\t\sum}'_k\,\frac{\d^2}{\d\p_k^2}\,a^s+\half \hbar^2\s a^{-2s}{\t\sum}'_k\frac{1}{\Psi_0(\p)}\,\frac{\d^2\s\Psi_0(\p)}{\d\p_k^2}\,a^s\no\\
    &&=-\half\hbar^2\s a^{-2s}{\t\sum}'_k\,\frac{\d^2}{\d\p_k^2}\,a^s+\half{\t\sum}'_{k,k^*}(\p_{k^*}-\p_k)^2\s a^{s-2}+\half\s m_0^2{\t\sum}'_k\p_k^2\,a^s\no\\
    &&\hskip8em+\s g_0{\t\sum}'_k\p_k^4\,a^s+\half\hbar^2{\t\sum}'_k{\cal F}_k(\p)\,a^s-E_0\;. \label{eH}\en
    Indeed, this latter equation implicitly determines the ground-state wave function $\Psi_0(\p)$!

    It is important to observe that no normal ordering applies to the terms in the Hamiltonian. Instead,
    local field operator products are determined by an operator product expansion \cite{IOP}. In addition, note that the ``counterterm'' $\half\hbar^2\Sigma'_k{\cal F}_k(\p)$ does {\it not} depend on any parameters of the model and
    specifically not on $g_0$. This is because the counterterm is really a counterterm
    for the {\it kinetic energy}. This fact follows because not only is $\H\s\Psi_0(\p)=0$, but then $\H^q\s\Psi_0(\p)=0$ for all integer $q\ge2$. While $[\Sigma'_k\d^2/\d\p_k^2]\s\Psi_0(\p)$ may be a square-integrable function, the expression  $[\Sigma'_k\d^{2}/\d\p_k^{2}]^q\s\Psi_0(\p)$ will surely not be square integrable for suitably large $q$. To ensure that $\Psi_0(\p)$
    is in the domain of $\H^q$, for all $q$, the derivative term and the counterterm must be considered together to satisfy domain requrements, hence
    our claim that the counterterm should be considered as a renormalization of the kinetic energy.

    Since the counterterm does not depend on the coupling constant, it follows that the counterterm remains even when
    $g_0\ra0$, which means that the interacting quantum field theory does {\it not} pass to the usual free quantum field theory as $g_0\ra0$, but instead it passes to what we have called a {\it pseudofree quantum field theory}. This behavior is not unknown; see \cite{IOP} and references therein. As a relevant example, consider the classical (Euclidean)
        action functional (\ref{e1}). Regarding the separate components of that expression,
        and assuming both $m_0>0$ and $g_0>0$, a multiplicative inequality \cite{russ,book} states that
          \bn \{g_0\tint \p(x)^4\,d^n\!x\}^{1/2}\le{\tilde C}\tint[(\nabla\p)(x)^2+m_0^2\p(x)^2]\,d^n\!x
          \label{e333}\en
          where ${\tilde C}= (4/3)\s [\s g_0^{1/2}\s m_0^{(n-4)/2}\s]$ when $n\le4$ (the renormalizable cases), and  ${\tilde C}=\infty$ when  $n\ge5$ (the nonrenormalizable cases), which in the latter  case means there are fields for which the integral on the left side of the inequality diverges
          while the integral on the right side is finite.
          In other words,  there are different free and pseudofree classical theories when $n\ge5$. Thus, for $n\ge5$, it is reasonable to assume that the free quantum field theory is also different from the pseudofree quantum field theory.
          Moreover, the quantum models developed in this letter with the unconventional counterterm provide
   viable candidates for those quantum theories normally classified as nonrenormalizable, and they do so
   in such a manner that in a perturbation analysis divergences do not arise because all the underlying measures are equivalent and not mutually singular! (A discussion of the divergence-free properties from a perturbation point of view appears in \cite{IOP}, an analysis that also determines the dependence
   of $Z$, $m_0^2$, and $g_0$ on the parameters $a$ and $N'$.) Since the unconventional counterterm conveys good properties to the nonrenormalizable models, it is natural to extend such good behavior to the traditionally renormalizable models $(n\le4)$ by using the unconventional counterterm for them as well,
   giving them an alternative version of quantization.
   {\it Thus, we are led to adopt the lattice regularized Hamiltonian $\H$ (\ref{eH}), including the counterterm, for all spacetime dimensions $n\ge2$.}

   Additionally, the lattice Hamiltonian (\ref{eH}) also determines the lattice Euclidean action including the unconventional counterterm, which is then given by
     \bn I=\half \Sigma_{k,k^*}(\p_{k^*}-\p_k)^2\,a^{n-2}+\half m_0^2\Sigma_k\p_k^2\,a^n+g_0\Sigma_k\p_k^4
     \,a^n+\half\Sigma_k \hbar^2{\cal F}_k(\p)\,a^n \label{eL}\en
     where in this expression $k^*$ refers to all $n$ nearest neighbors of the site
     $k$ in the positive direction. Although the lattice form of the counterterm involves averages over field-squared values
     in nearby spatial regions of the central site, it follows that the continuum limit of the counterterm
     is local in nature, as noted previously. It is noteworthy that preliminary Monte Carlo studies
     based on the lattice action (\ref{eL}) support a nontrivial
     behavior of the $\p_4^4$ model exhibiting a positive renormalized coupling constant \cite{stank}.\v

\section{Comparison to traditional \\quantization results}
  There already exist well-defined, conventional results for $\p^4_2$ and $\p^4_3$, and, yes, we are proposing alternative quantizations for these models. Unlike the usual approach, there are no divergences in perturbative expansions in the new formulation after operator product expansions are introduced for the local operators, and models like $g_{0}\s\p^4_3+g'_{0}\s\p^8_3$---a sum of superrenormalizable and nonrenormalizable models---exhibit reasonable properties. For $\p^4_4$, traditional methods lead to a renormalizable theory, but nonperturbative methods support a trivial (free) behavior. On the other hand, our procedures are expected to be divergence free and nontrivial \cite{stank}. The cases $\p^4_n$, $n\ge5$, are conventionally nonrenormalizable and require an infinite number of distinct counterterms. The success of the electroweak model suggests that nonrenormalizable models are effective field theories,  good for low energy questions with a few perturbative corrections.  However, some idealized nonrenormalizable models have actually been completely solved (see, e.g.,
  \cite{ULSC,ULFE}), and thus self-consistent solutions may be possible for other models as well. Indeed,
  the solution for scalar fields presented in \cite{ULSC,book} has been critical in developing the approach presented in this letter; likewise an extension to fermion fields, inspired by \cite{ULFE}, is currently under investigation.

  An important key to constructing self-consistent solutions for nonrenormalizable models is the realization that the zero-coupling limit of interacting solutions is different from the usual free theory, e.g., as follows from (\ref{e333}); this distinction already holds for the classical theory and thus for the quantum theory. Indeed, this behavior already arises for a {\it single particle} with the classical action
     $A=\tint\{\half[\s{\dot x}(t)^2-x(t)^2]-g\s x(t)^{-4}\s\}\,dt$.
  Accepting that fact opens the door for alternative counterterms that favor operator product expansions over normal ordering, and can lead to divergence-free formulations. Similar techniques also seem to be relevant for quantum gravity \cite{AQG-CC}.

  The inappropriateness of a perturbation analysis for the nonrenormalizable cases that follows from a difference
  between the pseudofree and free theory for such cases also eliminates the relevance of Landau poles for such
  problems. Those who believe the renormalization group is applicable are asked to consider the lattice action (\ref{eL}) with its specific form and the special dependence of its coefficients on the lattice spacing $a$. The coefficients in the counterterm (\ref{eF}) have been {\it designed} to lead to a nontrivial continuum limit. That special dependence on the lattice spacing would be hard to maintain throughout when taking the continuum limit via a renormalization group analysis.

\section{Acknowledgements}
  Thanks are extended to B. Bahr, G. Hegerfeldt,  S. Shabanov, C. Thorn, and R. Woodard for helpful comments.

\end{document}